\documentclass{emulateapj}

\shortauthors{Brammer and van Dokkum}
\shorttitle{}
\slugcomment{Accepted for publication in ApJL}

\newcommand{\MSOL}{\mbox{$\:{\rm M}_{\sun}$}} 
\newcommand{\zphot}{\mbox{$z_{\rm phot}$}} 
\newcommand{\zspec}{\mbox{$z_{\rm spec}$}} 
\newcommand{\ang}{\mbox{$\:$\AA}}

\begin{document}

\title{The Density and Spectral Energy Distributions of Red Galaxies at $z\sim3.7$}

\author{Gabriel B. Brammer\altaffilmark{1} and Pieter G. van
Dokkum\altaffilmark{1}} \email{brammer@astro.yale.edu}

\altaffiltext{1}{Department of Astronomy, Yale University,  New Haven,
CT, 06520-8101}

\begin{abstract}

We use the deep NIR imaging of the FIRES survey to investigate trends with
redshift of the properties of galaxies selected to have strong
Balmer/4000\ang\ breaks at $2<z<4.5$.  Analagous to the $J-K>1.3$ (AB) color
criterion designed to select red galaxies at $z>2$, we propose two color
criteria, $J-H>0.9$ and $H-K>0.9$, to select red galaxies in two redshift bins
at $2<z<3$ and $3<z<4.5$, respectively.  From the FIRES catalogs of the HDF-S
(4.7 arcmin$^2$) and MS 1054-03 (26.3 arcmin$^2$) fields, we find 18 galaxies
with $\left<z_\mathrm{phot}\right>=2.4$ that satisfy $J_s-H>0.9; H<23.4$ and
23 galaxies with $\left<z_\mathrm{phot}\right>=3.7$ that satisfy $H-K_s>0.9;
K_s<24.6$, where the flux limits are chosen to match the limiting rest-frame
luminosities at the different median redshifts of the two samples.  The space
densities of the $J_s-H$ and $H-K_s$ samples are $1.5\pm0.5\times10^{-4}$ and
$1.2\pm0.4\times10^{-4}\ \mathrm{Mpc}^{-3}$, respectively.  The rest-frame
$U-B$ colors of galaxies in both samples are similarly red (as expected from
the definition of the color criteria), but the rest-frame UV properties are
different:  galaxies in the higher-redshift $H-K_s$ selected sample have blue
NUV-optical colors and UV slopes similar to those of Lyman Break Galaxies,
while the $J_s-H$ galaxies are generally red over the entire wavelength range
observed.  Synthetic template fits indicate that the distinct rest-NUV
properties of the two samples are primarily a result of dust:  we find 
$\left<A_V\right>_{JH}=1$ mag and $\left<A_V\right>_{HK}=0.2$ mag.  The median
stellar mass determined from the template fits decreases by a factor of
$\sim5$ from $z=2.4$ to $3.7$, which, coupled with the fact that the space
density of such galaxies remains roughly constant, may imply that the stellar
mass density in red galaxies decreases by a similar factor over this
redshift range.

\end{abstract}


\section{Introduction}

Over the last few years, the study of galaxies with red rest-frame optical
colors has been extended to ever-increasing redshifts.  While the selection of
galaxies based on their rest-frame UV emission \citep[e.g. the Lyman break
technique;][]{steidel93} has enabled the detailed study of young galaxies at
high redshift for some time, it is only with the more recent advent of
efficient, large-scale detectors in the near-IR that large numbers of galaxies
with redder rest-frame colors have been discovered at early cosmic epochs
\citep{franx03,vd03}.  Red galaxies with typically very low UV fluxes make up
80\% of the mass contained in the most massive galaxies \citep{vd06} and 25-75\%
of the total mass in galaxies at $2<z<3$  \citep{papovich06, marchesini06}, and
thus provide critical constraints on theoretical models of galaxy formation and
evolution \citep{somerville04, nagamine05}.

A proven technique for selecting galaxies with red rest-frame optical colors is
the $J_s-K_s>1.3$ criterion \citep[$J-K_\mathrm{Vega}>2.3$;][]{franx03}, which
relies on the Balmer/4000\ang\ break redshifted into the $J$ band for redshifts
$z>2$.  Galaxies selected using this technique comprise a heterogeneous
population showing a broad range in dust properties, luminosity-weighted ages,
and star-formation rates \citep{forster04,labbe05,papovich06,kriek06b}. 

In this Letter, we extend the study of red galaxies to redshifts $z>3$, and we
compare the number and properties of red galaxies at $z\sim3.5$ to those at
$z\sim2.5$ using uniform color selection criteria based on the Balmer/4000\ang\
break redshifted into the $H$ and $J_s$ bands, respectively.  While there is
typically a tail of sources extending beyond $z>3.5$ in $J-K$ selected samples
\citep[e.g.][]{forster04}, there have been no systematic studies comparing the
numbers and properties of red galaxies at the extremes of the broad $J-K$
redshift selection window.  We adopt $H_0=70$ km/s, $\Omega_{\mathrm m} = 0.3$,
and $\Omega_\Lambda = 0.7$.  All magnitudes are given in the AB system.

\section{Color Selection of Red Galaxies}

\begin{figure}
\plotone{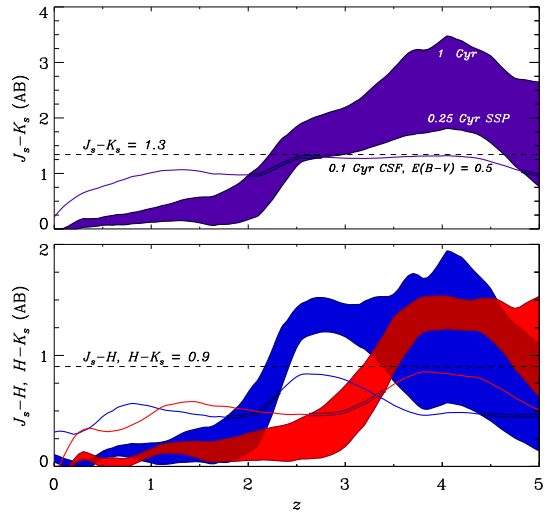}
\caption{\textit{Top}: Evolution of $J_s-K_s$ color with redshift.  The dashed
line shows the DRG criterion of \cite{franx03} designed to select
passively-evolving galaxies at $z>2$. \textit{Bottom}: $J_s-H$ (blue) and
$H-K_s$ (red) $vs.\ z$.  The dashed line indicates the selection criteria used
in this Letter, which are analogous to the DRG color selection but that divide
the DRG sample into two redshift bins.  The shaded bands indicate the range of
colors of dust-free \cite{bc03} passively evolving templates with ages between
0.25 and 1.0 Gyr.  The thin lines correspond to a template with a constant star
formation rate with age 0.1 Gyr and $E(B-V) = 0.5$, showing that moderately
reddened starbursts at low redshift are not expected to significantly
contaminate the $z>2$ NIR-selected galaxy samples.
\label{fig:colors}}
\end{figure}

The top panel of Fig.~\ref{fig:colors} shows the observed $J_s-K_s$ color of
\cite{bc03} template spectra over $0<z<5$.  At a given redshift, $J_s-K_s$ is
reddest for the oldest template because the strength of the Balmer/4000\ang\
break increases as a stellar population ages.  Based on the figure,
$J_s-K_s>1.3$ should select galaxies at $z>2$ that are either dominated by an
evolved stellar population or are highly reddened by dust---the so-called
``distant red galaxies'', or DRGs \citep{franx03}.  A number of DRGs have been
spectroscopically confirmed \citep{vd03,kriek06,kriek06b} at $z>2$, and large
photometric samples of DRGs are found to have $1.5\lesssim\zphot\lesssim3.5$
\citep{forster04,papovich06,quadri06}.

Because DRGs have a fairly broad redshift distribution, the rest-frame
properties of galaxies satisfying the color criterion to a given magnitude
limit in a particular selection band change with redshift in three important
ways. First, the limiting absolute magnitude in the selection band becomes
brighter with increasing redshift for a fixed survey depth, probing sharply
decreasing source densities at the bright end of the steep luminosity function.
Second, the rest wavelength of the selection band decreases with increasing
redshift, probing wavelengths where the scatter in $M/L_\lambda$ is large. 
Finally, the rest-frame color---essentially the type of galaxy
selected---changes with redshift because the Balmer/4000\ang\ break is narrow
in wavelength compared to the spacing between the $J_s$ and $K_s$ filters and
because the spectral slope is not the same on both sides of the break for galaxies
older than $\sim100$ Myr.  Together, these effects make it very difficult to
compare DRG properties at different redshifts.

The bottom panel of Fig.~\ref{fig:colors} demonstrates how splitting the $J-K$
criterion into two should divide DRGs into two redshift bins,
\begin{eqnarray} 
J_s-H>0.9\:&:&\:z\gtrsim2 \nonumber \\ 
H-K_s>0.9\:&:&\:z\gtrsim3,
\label{eq1}
\end{eqnarray}

\noindent mitigating the problems described above that prevent direct analysis
of the variation of red galaxy properties with redshift.  The wavelength
baselines of the two NIR colors are similar enough that the same color limit
can be used for both criteria to select identical rest-frame powerlaw spectral
slopes.  The criteria of Equation \ref{eq1} are adopted analagous to
$J_s-K_s>1.3$ to select against low-$z$ interlopers, while the selection
efficiency of the $J_s-H$ and $H-K_s$ criteria may be further enhanced by the
fact that those colors are steeper functions of redshift at the low-$z$
selection boundaries than $J_s-K_s$.

Alternatively, one could select samples of galaxies based on their rest-frame
properties estimated using distances determined from the galaxies' photometric
redshifts.  We note, however, that the $\zphot$--$\zspec$ calibration is poorly
determined at $z>3$, and therefore relying on photometric redshifts alone
introduces large uncertainties in the analysis at high redshift.  Furthermore,
by working in the observed frame, the results can be easily verified by others
independent of photometric redshift or analysis techniques.

\section{Data \label{sec_data}}

We use the deep optical+NIR photometry of the FIRES survey \citep{franx00} to
select red galaxies at $z>2$  based on the two NIR color criteria described
above.  Details of the data reduction and $K$-selected source catalogs of the
two fields of the survey, HDF-S (4.7 arcmin$^2$) and MS 1054-03 (26.3
arcmin$^2$),  can be found in \cite{labbe03} and \cite{forster06},
respectively.  Briefly, the combined catalog of the two fields contains HST
$U_{300}B_{450}$ (HDFS-S), $V_{606}I_{814}$ (both fields), ground-based $UBV$
(MS 1054-03) and ISAAC-$J_sHK_s$ (both fields) photometry, along with
photometric redshifts determined following the procedure described by
\cite{rudnick01,rudnick03}.  Details of the accuracy of the photometric
redshifts in these fields are given in \cite{labbe03} and \cite{forster06}. 
The combined catalog depth is limited by the shallower MS 1054-03
observations.  At $K_s=24.6$, the combined catalog is $\sim90\%$ complete and
sources in the MS 1054-03 catalog have $\left(S/N\right)_K=5-7$.  The FIRES
dataset provides a unique opportunity to select red galaxies at $z>3.5$---even
the brightest of which should still be quite faint in $K$---since  there are no
other currently available surveys of comparable depth in all three $JHK$ NIR
filters.

\section{Densities \label{sec_density}}

\begin{figure}  
\plotone{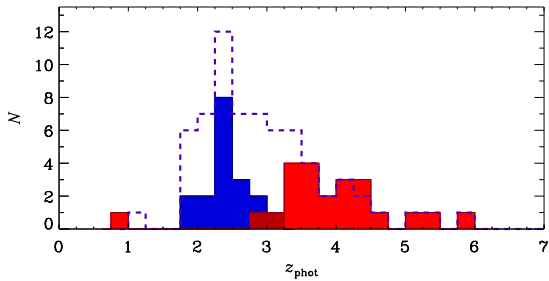}
\caption{Photometric redshift distribution of galaxies in the FIRES fields with
$J_s-H>0.9; H<23.4$ (blue) and $H-K_s>0.9; K_s<24.6$ (red). Although the samples show
a small overlap in redshift space, no galaxy meets both selection criteria.  For
comparison, the dotted line shows the redshift distribution of galaxies satisfying
$J_s-K_s>1.3$ and $K_s<24.6$.
\label{fig:zhist}}
\end{figure}

From the combined FIRES catalog we find 18 sources that satisfy
$J_s-H>0.9\:;\:H<23.4$  and 23 sources with $H-K_s>0.9\:;\:K_s<24.6$.  The
limiting magnitude in $H$ was chosen such that the rest-frame limiting magnitude
redward of the Balmer/4000\ang\ break is the same for both samples, and its
value was determined based on the ratio of the luminosity distances at the
median redshifts of the two samples.  The number of selected sources correspond
to combined (HDF-S-only) surface densities and Poisson errors of
$0.58\pm0.18\;(0.63\pm0.37)$ and $0.74\pm0.19\;(0.42\pm0.30)$ arcmin$^{-2}$ for
the $J_s-H$ and $H-K_s$ selected samples, respectively.  The photometric
redshift distributions of the NIR selected sources are shown in
Fig.~\ref{fig:zhist}.  The $J_s-H$ sample has $\zphot=2.4\pm0.3$ (1-$\sigma$
range) and 16/18 galaxies also satisfy $J_s-K_s>1.3$.  For the $H-K_s$ sample,
$\zphot=3.9\pm1.0$ and 15/23 galaxies satisfy $J_s-K_s>1.3$.

If we consider comoving volumes bounded by tophat redshift distributions of
$2<\zphot<3$ and $3<\zphot<4.5$, the area-weighted space densities and
associated Poisson errors of the $J_s-H$ and $H-K_s$ samples are
$1.5\pm0.5\times10^{-4}$ and $1.2\pm0.4\times10^{-4}\ \mathrm{Mpc}^{-3}$,
respectively.  Therefore, the space density of a flux-limited sample of red
galaxies remains constant within the errors from $\left<z\right>_{\rm
JH}\sim2.4$ to $\left<z\right>_{\rm HK}\sim3.7$.  We note that the total survey
area discussed here is relatively small and that our result is likely subject to
cosmic variance, especially in light of recent studies that show that red
galaxies at $z>2$ are strongly clustered \citep{daddi03,quadri06}.

\section{Rest Frame SEDs \label{sec_sed_fits}}

\begin{figure} 
\plotone{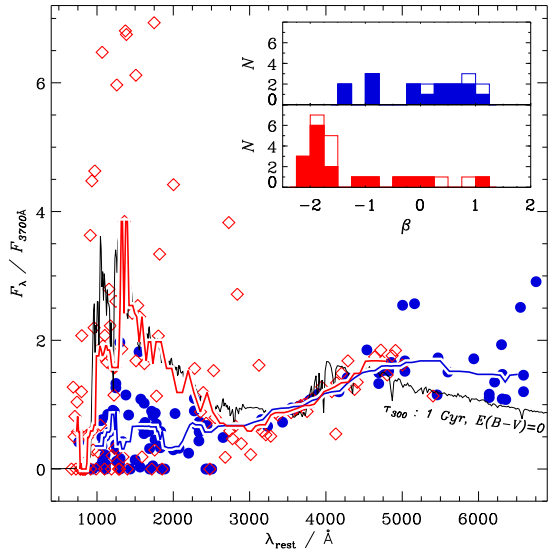}
\caption{Rest-frame SEDs of the $J_s-H$ (blue circles) and $H-K_s$ (red
diamonds) galaxy samples limited to $2<\zphot<3$ and $3<\zphot<4.5$,
respectively.  The SEDs are normalized to the flux at $\lambda_{\rm rest} =
3700$\ang\, determined from a linear interpolation between the two nearest
filters in the rest frame.  The solid red and blue lines indicate a running
median of the 10 neighboring points for the two samples.  An unreddened 1 Gyr
old $\tau_{300}$ template is shown for reference.  The similarity of the SED
slopes at the Balmer/4000\ang\ break demonstrates how the two color criteria
are matched to select similar rest-frame spectral shapes at two different
redshifts.  $Inset$: distribution of the synthetic rest-frame UV spectral
slope, $\beta$.  The thin histograms correspond to the entire samples, while
the filled histograms are for sources that fall within the $2<\zphot<3$ and
$3<\zphot<4.5$ comparison ranges.  With few exceptions, the higher redshift
galaxies in the $H-K_s$ sample have UV-optical colors that are significantly
bluer than those of the $J_s-H$ selected galaxies at lower redshift.
\label{fig:seds}} 
\end{figure}

Athough there is no appreciable change in the \textit{number} of galaxies
selected to have strong Balmer/4000\ang\ breaks from $z\sim2.4$ to $z\sim3.7$,
their properties are quite different.  Fig.~\ref{fig:seds} shows the spectral
energy distributions (SEDs) of the two galaxy samples shifted to the
rest-frame.  The $J_s-H$ galaxies typically have red rest-frame UV-optical
colors and a fairly flat rest-UV spectrum, consistent with previous studies of
DRGs in the redshift range $2<z<3$ \citep{forster04}.  In contrast, the
higher-redshift $H-K_s$ galaxies generally have blue UV-optical colors.  The
spectral slopes of the two samples through the Balmer/4000\ang\ break are
similar for the two samples, confirming that the NIR color criteria proposed
here select galaxies with similar $\left(U-B\right)_{\rm rest}$ colors over
$2<z<4.5$.

We quantify the spectral shapes of the SEDs using the rest-frame UV power-law
slope, $F_\lambda\propto\lambda^\beta$ \citep{calzetti94}, measured from a
best-fit \cite{bc03} template with an exponentially decaying star formation
rate ($\tau=300$ Myr) and solar metallicity.  The template fits to the
broadband photometry hold the redshift fixed to the catalog \zphot\ values,
allow ages between 0.1 Myr and the age of the universe at \zphot\, and allow
$A_V=0-3$ mag following the extinction law of \cite{calzetti00}.  Corrections
for Ly$\alpha$ forest absorption are applied following \cite{madau95}.  The
distributions of $\beta$ for the two galaxy samples are shown in
Fig.~\ref{fig:seds}.  The distribution is quite flat for the $J_s-H$
sample---similar to the distribution seen for a large sample of massive DRGs
by \cite{vd06}.  In contrast, the distribution of $\beta$ for the $H-K_s$
sample shows a peak at $\beta\sim-2$, values similar to those found by
\cite{adelberger00} for UV-selected galaxies and to the sample of massive
Lyman break galaxies (LBGs) discussed by \cite{vd06}.  

Of the 12 galaxies in the $H-K$ sample that have \zphot\ in the range
$2.7<z<3.4$ or $3.9<z<4.5$, 10 have synthetic $U_nG{\cal R}I$ colors
integrated from the best-fit templates that satisfy the $U$ or $G$ dropout LBG
color criteria at those redshifts proposed by \cite{steidel99}.  However, 8 of
those 10 galaxies with LBG colors have ${\cal R}>25.5$, too faint to be
included in typical spectroscopic samples of LBGs.  

\section{Discussion}

We have shown that we can efficiently select $z\sim3.7$ galaxies in the near-IR
with the simple color criterion $H-K>0.9$.  The  samples described here
indicate that the rest-frame UV-optical SEDs of galaxies selected to have
strong Balmer/4000\ang\ breaks are significantly different at $z\sim2.4$ and
$z\sim3.7$:  galaxies in the higher redshift sample have a median NUV/optical
flux ratio $2-4$ times greater than that of the galaxies at $z\sim2.4$. 
Finding evolution in the properties of galaxies over the $\sim1$ Gyr between
$z=2.4-3.7$, which at the distant end is only 1.7 Gyr after the Big Bang, is in
itself not surprising, as the spectral evolution is rapid at these redshifts
for galaxies with a broad range of formation redshifts and subsequent star
formation histories.  What is interesting is that the galaxies in both samples
were selected to have similarly strong Balmer/4000\ang\ breaks, indicating the
presence of an evolved stellar population that itself would not be a likely
source for the strong NUV component of the $z\sim3.7$ SEDs.  A two-burst model
whose optical SED is dominated by an evolved stellar population but that also
contains a young component that supplies the UV flux could explain the
$z\sim3.7$ SEDs.  Though two-component model fits are beyond the scope of this
Letter, there have been other observations of red galaxies at high-$z$
suggesting that composite populations are appropriate, both from SED modelling
\citep{yan04} and from observations of distinct UV and optical galaxy
morphologies \citep{toft05} of such galaxies.  

Our $\tau_{300}$ template fits suggest that dust is the primary source of the
difference in spectral shape shown in Fig.\ref{fig:seds}.  Both samples have a
median template age of $\sim1$ Gyr, while the median dust extinction decreases
from $A_V=1$ to 0.2 mag going from the $J_s-H$ to the $H-K_s$ sample.  In
general one expects the dust content of galaxies, along with the metal content,
to increase with time, qualitatively consistent with the analysis presented
here.  It is difficult to quantitatively compare our results to theoretical
models of galaxy formation since such models typically rely on ad hoc treatments
of dust absorption \citep[e.g.][]{croton06}.  Interestingly, our results may be
qualitatively consistent with the redshift distribution of
submillimeter-selected galaxies.  Although still quite uncertain, there is
evidence that the number of very dusty, luminous galaxies drops significantly
from $z\sim2.4$ to $z\sim3.7$ \citep{chapman05}.

The normalization of our population synthesis fits provides a rough estimate of
the stellar mass of each galaxy in the sample.  With our photometry sampling
only the rest-frame UV-optical light of the galaxies in our sample at $z>2$, the
stellar mass fits are uncertain to factors of $>2$ due to uncertainties in
$\zphot$ and in the IMF and to model degeneracies between age, dust,
metallicity, and star formation history.  With those caveats in mind, our
template fits imply that the median stellar mass of red galaxies decreases by a
factor of $\sim5$ from $z\sim2.4$ to $z\sim3.7$.  While the uncertainties are
large for the masses of individual galaxies, a Mann-Whitney test on the
distribution of masses suggests that the difference in the median masses of the
two samples is significant at the 99\% confidence level.  \cite{erb06} observe a
similar trend in the mass of LBGs:  the average dynamical and stellar masses of
$z\sim3$ LBGs are a factor of 2 smaller than for LBGs at $z\sim2$.  We note here
that the $H-K$ galaxies are not necessarily direct progenitors of the $J-H$
galaxies, just as LBGs at $z=4$ are not necessarily progenitors of LBGs at
$z=2$.  Many of the $H-K$ galaxies could fade below our magnitude limit after
$\sim1$ Gyr; conversely, many of the $J-H$ galaxies could have been much bluer 1
Gyr previously. 

Since the space densities of the $J_s-H$ and $H-K_s$ samples are statistically
equivalent, a decrease in the median stellar mass in the higher redshift sample,
if real, would indicate a decrease in the stellar mass density of galaxies
with evolved stellar populations.  This may be consistent with the results of \cite{kriek06b}, who find that the
ages of apparently passive galaxies at $z\sim2.3$ (which make up 45\% of their
$K$-selected sample) are typically $\lesssim1$ Gyr. It seems unlikely that many
of the $z>3$ progenitors of these galaxies were already  ``red and dead''. 
On the other hand, this result may be difficult to reconcile with the existence
of a significant population of $M_*>10^{11}\MSOL$ galaxies at $z>6$, as may be
implied by the source described by \cite{mobasher05}.  Furthermore, the best fit
\cite{bc03} $z=6.5$ template of the \cite{mobasher05} source has $\beta\sim2$
and would be an outlier even at $z\sim3.7$ compared to the sources in our $H-K$
sample.

Clearly, much further work is required to fully understand how the properties
of red galaxies change with time at $z>2$.  The differences shown in
Fig.~\ref{fig:seds} are model independent, to the extent that the photometric
redshifts approximate the true redshifts, but the results above based on
template fits are quite uncertain. The addition of $Spitzer$ photometry would
better constrain the stellar mass contained in these galaxies
\citep{wuyts06}.  The brightest galaxies in the $H-K_s$ sample may be within
reach of NIR spectroscopy that could precisely determine redshifts and model
stellar populations based on the prominent Balmer/4000\ang\ breaks, as has
been done quite effectively for DRGs at $z\sim2.5$ \citep{kriek06}.  Finally
we note that the UKIDSS Ultra Deep Survey is planned to be $\sim$0.2 mag
deeper than our adopted $K_s$ limit over an area $\sim90$ times that of
FIRES, which would vastly increase the sample sizes of galaxies selected as
described here.  

\begin{acknowledgements}

We thank Ivo Labb{\'e} and Natascha F{\"o}rster Schreiber for providing the FIRES
data to the community, Ryan Quadri for discussion and comments, and the referee,
Michael Rowan-Robinson, for his suggestions that improved this Letter.  Support
from National Science Foundation grant NSF CAREER AST 04-49678 is gratefully
acknowledged. 

\end{acknowledgements}


\begin{thebibliography}{}

\bibitem[Adelberger \& Steidel(2000)]{adelberger00} Adelberger, 
K.~L., \& Steidel, C.~C.\ 2000, \apj, 544, 218 

\bibitem[Bruzual \& Charlot(2003)]{bc03} Bruzual, G.~\&~Charlot, S.\ 2003,
\mnras, 344, 1000

\bibitem[Calzetti et~al.(1994)]{calzetti94} Calzetti, D., Kinney, 
A.~L., \& Storchi-Bergmann, T.\ 1994, \apj, 429, 582 

\bibitem[Calzetti et~al.(2000)]{calzetti00} Calzetti, D., Armus, 
L., Bohlin, R.~C., Kinney, A.~L., Koornneef, J., \& Storchi-Bergmann, T.\ 
2000, \apj, 533, 682 

\bibitem[Chapman et~al.(2005)]{chapman05} Chapman, S.~C., Blain, 
A.~W., Smail, I., \& Ivison, R.~J.\ 2005, \apj, 622, 772 

\bibitem[Croton et~al.(2006)]{croton06} Croton, D.~J., et~al.\ 
2006, \mnras, 365, 11 

\bibitem[Daddi et~al.(2003)]{daddi03} Daddi, E., et~al.\ 2003, 
\apj, 588, 50 

\bibitem[Erb et~al.(2006)]{erb06} Erb, D.~K., Steidel, C.~C.,  Shapley, A.~E.,
Pettini, M., Reddy, N.~A., \& Adelberger, K.~L.\ 2006,  \apj, 646, 107

\bibitem[F\"orster-Schreiber et~al.(2004)]{forster04} F\"orster-Shreiber, N.~M.,
et~al.\ 2004, \apj, 616, 40

\bibitem[F\"orster Schreiber et~al.(2006)]{forster06} F\"orster Shreiber, N.~M.,
et~al.\ 2006, \aj, 131, 1891

\bibitem[Franx et~al.(2000)]{franx00} Franx, M., et~al.\ 2000, 
The Messenger, 99, 20 

\bibitem[Franx et~al.(2003)]{franx03} Franx, M., et~al.\ 2003, \apjl, 587, L79

\bibitem[Kriek et~al.(2006a)]{kriek06} Kriek, M., et~al.\ 2006, \apj, 645, 44

\bibitem[Kriek et~al.(2006b)]{kriek06b} Kriek, M., et~al.\ 2006, \apjl, 649, L71

\bibitem[Labb{\'e} et~al.(2003)]{labbe03} Labb{\'e}, I., et~al.\ 2003, \aj, 125,
1107

\bibitem[Labb{\'e} et~al.(2005)]{labbe05} Labb{\'e}, I., et~al.\ 2005, \apjl,
624, L81 

\bibitem[Madau(1995)]{madau95} Madau, P. 1995, \apj, 441, 18

\bibitem[Marchesini et~al.(2006)]{marchesini06} Marchesini, D., et~al.\ 2006,
\apj, submitted (astro-ph/0610484)

\bibitem[Mobasher et~al.(2005)]{mobasher05} Mobasher, B., et~al.\ 2005, \apj,
635, 832

\bibitem[Nagamine et~al.(2005)]{nagamine05} Nagamine, K., Cen, R., Hernquist,
L., Ostriker, J.~P. \& Springel, V.\ 2005, \apj, 618, 23

\bibitem[Papovich et~al.(2006)]{papovich06} Papovich, C., et~al.\ 2006, \apj,
640, 92

\bibitem[Quadri et~al.(2006)]{quadri06} Quadri, R., et~al.\  2006, \apj,
submitted (astro-ph/0606330)

\bibitem[Rudnick et~al.(2001)]{rudnick01} Rudnick, G. et~al.\ 2001, \aj, 122,
2205

\bibitem[Rudnick et~al.(2003)]{rudnick03} Rudnick, G. et~al.\ 2003, \apj, 599,
847

\bibitem[Somerville et~al.(2004)]{somerville04} Somerville, R.~S. et~al.\ 2004,
\apj, 600, L135

\bibitem[Steidel \& Hamilton(1993)]{steidel93} Steidel, C.~C. \& Hamilton, D.\
1993, \aj, 105, 2017

\bibitem[Steidel et~al.(1999)]{steidel99} Steidel, C.~C., Adelberger, K.~L.,
Giavalisco, M., Dickinson, M. \& Pettini, M.\ 1999, \apj,519,1

\bibitem[Steidel et~al.(2004)]{steidel04} Steidel, C.~C., 
Shapley, A.~E., Pettini, M., Adelberger, K.~L., Erb, D.~K., Reddy, N.~A., 
\& Hunt, M.~P.\ 2004, \apj, 604, 534 

\bibitem[Toft et~al.(2005)]{toft05} Toft, S., van Dokkum, P., 
Franx, M., Thompson, R.~I., Illingworth, G.~D., Bouwens, R.~J., \& Kriek, 
M.\ 2005, \apjl, 624, L9 

\bibitem[van Dokkum et~al.(2003)]{vd03} van Dokkum, P.~G., et~al.\ 2003, \apjl,
587, L83

\bibitem[van Dokkum et~al.(2006)]{vd06} van Dokkum, P.~G., et~al.\ 2006, \apjl,
638, L59

\bibitem[Wuyts et~al.(2006)]{wuyts06} Wuyts, S., et~al.\ 2006, 
\apj, in press (astro-ph/0609548)

\bibitem[Yan et~al.(2004)]{yan04} Yan, H., et~al.\ 2004, 
\apj, 616, 63 

\end{thebibliography}
\end{document}